\documentclass[letter, twocolumn]{jpsj3} 
\usepackage{epstopdf}

\title{Magnetotransport Studies of EuFe$_2$As$_2$: the Influence of the Eu$^{2+}$ Magnetic Moments}

\author{Taichi \textsc{Terashima}$^{1, 4}$, Nobuyuki \textsc{Kurita}$^{1, 4}$, Akiko \textsc{Kikkawa}$^{2}$, Hiroyuki S. \textsc{Suzuki}$^{2}$, Takehiko \textsc{Matsumoto}$^{2}$, Keizo \textsc{Murata}$^{3}$, and Shinya \textsc{Uji}$^{1, 4}$}

\inst{$^{1}$National Institute for Materials Science, Tsukuba, Ibaraki 305-0003\\
$^{2}$National Institute for Materials Science, Tsukuba, Ibaraki 305-0047\\
$^{3}$Division of Molecular Materials Science, Graduate School of Science, Osaka City University, Osaka 558-8585\\
$^{4}$JST, Transformative Research-Project on Iron Pnictides (TRIP), Chiyoda, Tokyo 102-0075}

\abst{We report resistivity $\rho$ and Hall effect measurements on EuFe$_2$As$_2$ at ambient pressure and 28 kbar and magnetization measurements at ambient pressure.  We analyze the temperature and magnetic-field dependence of $\rho$ and the Hall effect using a molecular-field theory for magnetoresistance and an empirical formula for the anomalous Hall effect and find that electron scattering due to the Eu$^{2+}$ local moments plays only a minor role in determining electronic transport properties of EuFe$_2$As$_2$.}

\kword{iron pnictides, pressure-induced superconductivity, resistivity, Hall effect}

\begin{document}
\maketitle

\def\degree{\kern-.2em\r{}\kern-.3em}

Since the discovery of superconductivity (SC) in LaFeAsO$_{1-x}$F$_x$ at a transition temperature $T_c$ = 26 K by Kamihara \textit{et al}.\cite{Kamihara08JACS}, extensive studies have revealed that SC can be induced in layered iron pnictides by a variety of tuning parameters: i.e., carrier doping, isovalent substitution (chemical pressure), and pressure $P$.\cite{Ishida09JPSJ_review}  Obviously, the key is not the introduction of carriers but the suppression of the antiferromagnetic (AFM) order of the Fe magnetic moments.  Generally, suppression of antiferromagnetism in metallic systems may cause non-Fermi liquid behavior.  Indeed, linear-in-temperature $T$ resistivity $\rho$, suggestive of strong electron scattering due to critical or nearly critical two-dimensional AFM spin fluctuations,\cite{Moriya00AdvPhys} has been observed in various optimally tuned iron pnictide superconductors,\cite{LiuR08PRL, Ning09JPSJ, Gooch10PRB, Jiang09JPCM} and in some cases NMR data appear to suggest the existence of a magnetic quantum critical point (QCP).\cite{Ning10PRL, Nakai10condmat}  Although a QCP in the iron pnictides is still highly debatable,\cite{Luetkens09nmat, Rullier-Albenque09PRL} further research into this issue is vital to determine relevance of spin fluctuations and quantum criticality to the iron pnictide high-$T_c$ SC.

It was previously pointed out that the temperature exponent of $\rho$ in metals near an AFM QCP is sensitive to disorder.\cite{Rosch99PRL}   In this regard, pressure tuning may better be suited than doping or substitution for electronic transport studies of the quantum criticality.  In a previous paper,\cite{Terashima09JPSJ_EFA} we have established that EuFe$_2$As$_2$ exhibits bulk SC at $P \sim$26 kbar.  This critical pressure for the appearance of bulk SC is lowest among the 122 compounds under good hydrostatic conditions\cite{Alireza09JPCM, Matsubayashi09JPSJ, Yamazaki10PRB, Yu09PRB} and is easily accessible by piston-cylinder high-pressure cells.  EuFe$_2$As$_2$ therefore seems the best system to study spin fluctuations and possible quantum criticality through transport measurements unless electron scattering due to the Eu$^{2+}$ localized magnetic moments masks signs of Fe spin fluctuations.  In this paper, based on magnetotransport and magnetization data, we show that the Eu$^{2+}$ moments play only a minor role in determining the electronic transport properties of EuFe$_2$As$_2$ and hence that the possible quantum criticality can be studied through more systematic transport studies of EuFe$_2$As$_2$ in future. 

EuFe$_2$As$_2$ exhibits two phase transitions, at $T_o\sim$ 190 K and $T_N\sim$ 19 K, at ambient $P$.\cite{Raffius93JPCS, Tegel08JPCM, Ren08PRB, Jeevan08PRB}  The transition at $T_o$ is a combined structural and magnetic transition, similar to those in other 122 compounds: the crystal structure changes from tetragonal to orthorhombic and the Fe moments order antiferromagnetically.  The magnetic structure of the Fe sublattice is the same as those in other 122 compunds.\cite{Xiao09PRB, Herrero-Martin09PRB}  The transition at $T_N$ is due to the AFM ordering of the Eu$^{2+}$ moments.  The magnetic structure of the Eu sublattice is composed of ferromagnetic Eu$^{2+}$ layers stacking antiferromagnetically along the $c$ axis.\cite{Xiao09PRB, Herrero-Martin09PRB}  The Eu$^{2+}$ moments are 6.8 $\mu_B$/Eu, aligned along the $a$ axis.  The interlayer AFM coupling of the Eu$^{2+}$ moments is rather weak: even below $T_N$, the paramagnetic state (the field-induced ferromagnetic state) of the Eu sublattice is easily reached by the application of $\sim$1 or 2 T in the $ab$-plane or along the $c$-axis, respectively.\cite{Terashima09JPSJ_EFA,Terashima09PhysicaC, Jiang09NJP, Xiao10PRB}  A $T-P$ phase diagram has been determined from $\rho$ measurements:\cite{Miclea09PRB, Kurita10SCES} while $T_o$ decreases with $P$ and is not detected above $P_c$ = 25--27 kbar (depending on crystal quality), $T_N$ is nearly $P$-independent.  The SC is observed below $T_c \sim$ 30 K near $P_c$. 

\begin{figure*}[tb]
\begin{center}
\includegraphics[width=16cm]{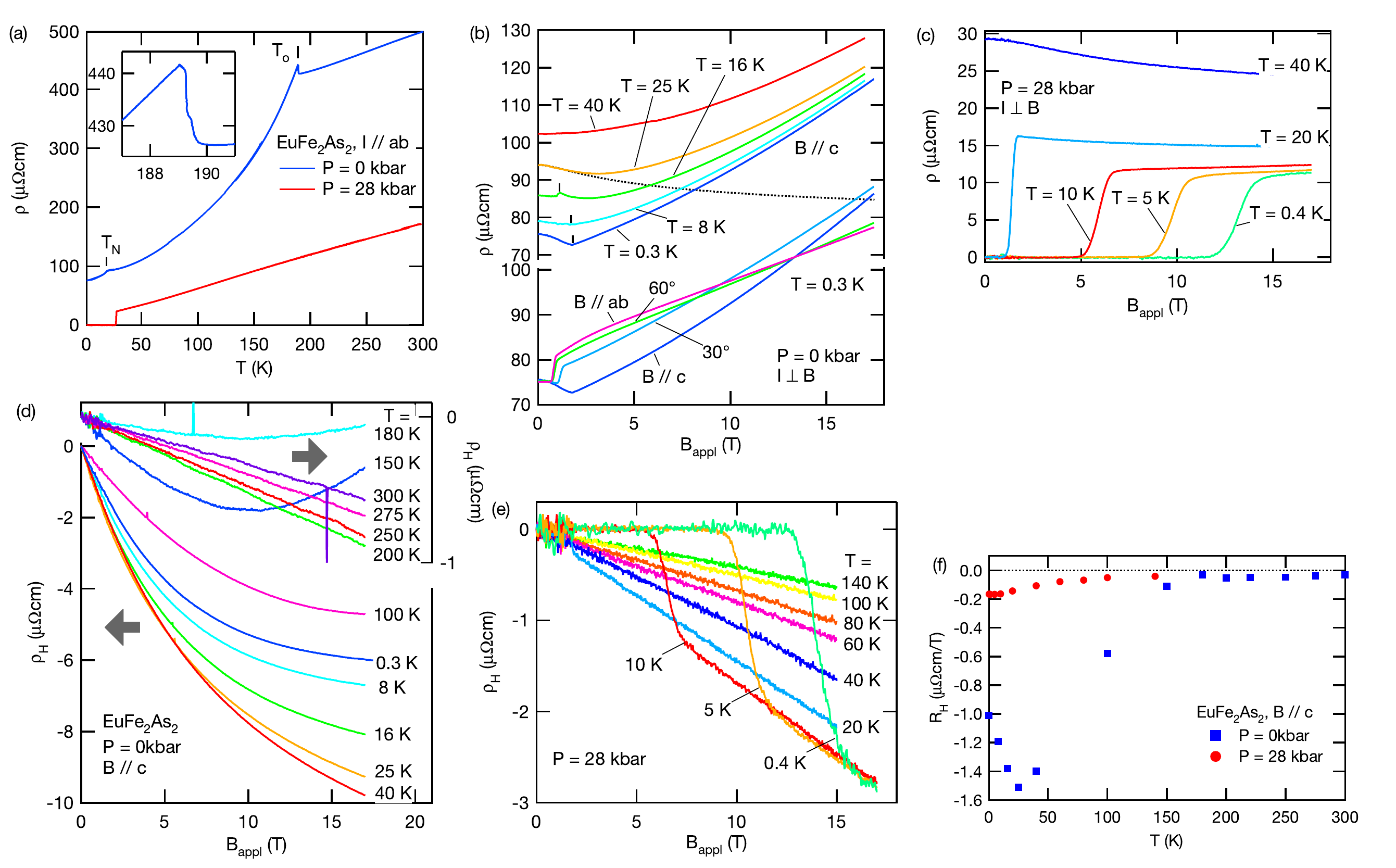}
\end{center}
\caption{(Color online) (a) Resistivity $\rho$ in EuFe$_2$As$_2$ as a function of $T$ at $P$ = 0 kbar and 28 kbar.  (b) and (c) $\rho$ as a function of $B_{appl}$. The dotted curve drawn for $T$ = 25 K in (b) is calculated from a molecular-field model (see text).  (d) and (e) Hall resistivity $\rho_H$  as a function of $B_{appl}$. (f) Hall coefficient $R_H$ as a function of $T$.  When $\rho_H$ is nonlinear, $R_H$ is defined as d$\rho_H$/d$B_{appl}$ at $B_{appl}$ = 0.}
\label{transport}
\end{figure*}

A single-crystal ingot of EuFe$_2$As$_2$ was grown by the Bridgman method from a stoichiometric mixture of the constituent elements.  Resistivity $\rho$ and Hall effect measurements were performed on thin (8 and 39 $\mu$m thick) samples exfoliated from the ingot with a usual six-contacts configuration using a low-frequency ac current ($I$ = 0.3 mA, $f \sim$17 Hz) applied in the $ab$ plane.  A $^3$He refrigerator and a superconducting magnet were used for the measurements.  For measurements at $P$ = 28 kbar, a clamped piston-cylinder cell was used, and the pressure-transmitting medium was Daphne7474 (Idemitsu Kosan Co., Ltd., Tokyo).\cite{Murata08RSI}  The magnetization measurements were performed on an approximately 1mm$^3$ sample using a SQUID magnetometer.  Note that we basically refer to the applied field $B_{appl}$, which may be different from the field $B$ inside a sample by $\sim$0.9 T, because of the large saturation moment of Eu$^{2+}$.  Also note that we use the convention $B$ = $\mu_0H$ + $M$, and hence that the magnetization $M$ is measured in the units of Tesla.

Figure~\ref{transport}(a) shows $\rho$ at $P$ = 0 and 28 kbar as a function of $T$.  At $P$ = 0 kbar, the two transitions at $T_o$ = 189 K and $T_N$ = 19 K are clearly observed.  A close examination of the anomaly at $T_o$ (inset) seems to suggest a two-step transition similar to those found in Ba(Fe$_{1-x}$Co$_x$)$_2$As$_2$.\cite{Chu09PRB}  The residual resistivity ratio $\rho$(300 K) / $\rho$(4 K) is 6.5.  At $P$ = 28 kbar, a nearly $T$-linear dependence is observed from RT down to $T_c$ = 28 K (midpoint), as is consistent with previous reports.\cite{Terashima09JPSJ_EFA, Kurita10SCES}  Figure~\ref{transport}(b) shows the field dependence of $\rho$ at $P$ = 0 kbar for various temperatures and also for various field directions (lower part).  For $B \parallel c$, the $T$ = 0.3 K curve shows a cusp at $B_{appl}$ = 1.8 T, which indicates an entrance into the paramagnetic (field-induced ferromagnetic) state of the Eu$^{2+}$ moments.  Above this transition field, $\rho$ increases rapidly, which is attributed to usual positive transverse magnetoresistance (MR) due to cyclotron motion of conduction carriers.  As $T$ increases, the anomaly shifts to lower fields, and it appears as a step-like increase at $T$ = 16 K.  At $T$ = 25 K ($> T_N$), no transition appears, and a negative MR at low fields is ascribable to suppression by fields of electron scattering due to Eu$^{2+}$.  At 40 K and higher temperatures below $T_o$, a monotonic positive MR is observed.  Above $T_o$, MR is practically zero.  The lower part of Fig.~\ref{transport}(b) shows that as $B_{appl}$ is tilted toward $ab$ the anomaly changes into a jump, and that the transition field decreases.  In a previous study,\cite{Jiang09NJP} the anomaly appeared as a cusp for both $B \parallel c$ and $B \parallel ab$ at $T$ = 2 K.  The origin of this difference is not clear.  Figure~\ref{transport}(c) shows the field dependence of $\rho$ at $P$ = 28 kbar.  In contrast to the $P$ = 0 kbar data, positive MR due to the cyclotron motion is not observed until 10 K as $T$ is lowered.  The origin of this contrast can be ascribed to the difference in the state of the Fe moments between $P$ = 0 and 28 kbar.  At $P$ = 28 kbar, the Fe moments do not order down to zero temperature so that their spin fluctuations make a large contribution to the resistivity, which can be suppressed by magnetic fields, leading to the negative MR in a wide $T$ range.

Figure~\ref{transport}(d) shows Hall resistivity $\rho_H$ at $P$ = 0 kbar as a function of $B_{appl}$.  While $\rho_H$ is linear in $B_{appl}$ above $T_o$, it exhibits pronounced nonlinearity below $T_o$.  The temperature dependence of the Hall coefficient $R_H$ is shown in Fig.~\ref{transport}(f), where we have defined $R_H$ as d$\rho_H$/d$B_{appl}$ at $B_{appl}$ = 0 when $\rho_H$ is nonlinear.  The obtained $R_H$($T$) is qualitatively or semi-quantitatively similar to previous polycrystal data\cite{Ren08PRB} and other 122 compounds data.\cite{Ni08PRB, Chen08PRB, Ronning08JPCM, Rullier-Albenque09PRL, Fang09PRB}  At RT $R_H$ is -0.033 $\mu\Omega$cm/T, which corresponds to 0.87 electron/Fe if a single-carrier model is assumed.  The magnitude of the Hall coefficient $|R_H|$ increases steeply below $T_o$, indicating the destruction of substantial part of the Fermi surface due to the AFM order of the Fe moments.  The maximum of $|R_H|$ at $T$ = 25 K ($R_H$ = -1.5$\mu\Omega$cm/T) corresponds to 0.019 electron/Fe in a single-carrier model.  Figure~\ref{transport}(e) shows $\rho_H$ at $P$ = 28 kbar as a function of $B_{appl}$ for various temperatures.  $\rho_H$ is linear in $B_{appl}$ for all the measured temperatures.  At $T$ = 140 K (the highest measured temperature), $R_H$ is -0.042 $\mu\Omega$cm/T, similar to the RT value at $P$ = 0 kbar.  As $T$ is lowered, $|R_H|$ increases gradually, reaching $R_H$ = -0.17 $\mu\Omega$cm/T at low temperatures, which corresponds to 0.17 electron/Fe in a single-carrier model.  A similar enhancement of $|R_H|$ at low $T$ in the paramagnetic state of the Fe moments has been observed in other optimally or nearly optimally tuned 1111 and 122 compounds.\cite{LiuR08PRL, Rullier-Albenque09PRL, Fang09PRB, Kasahara10PRB}

\begin{figure}[tb]
\begin{center}
\includegraphics[width=8cm]{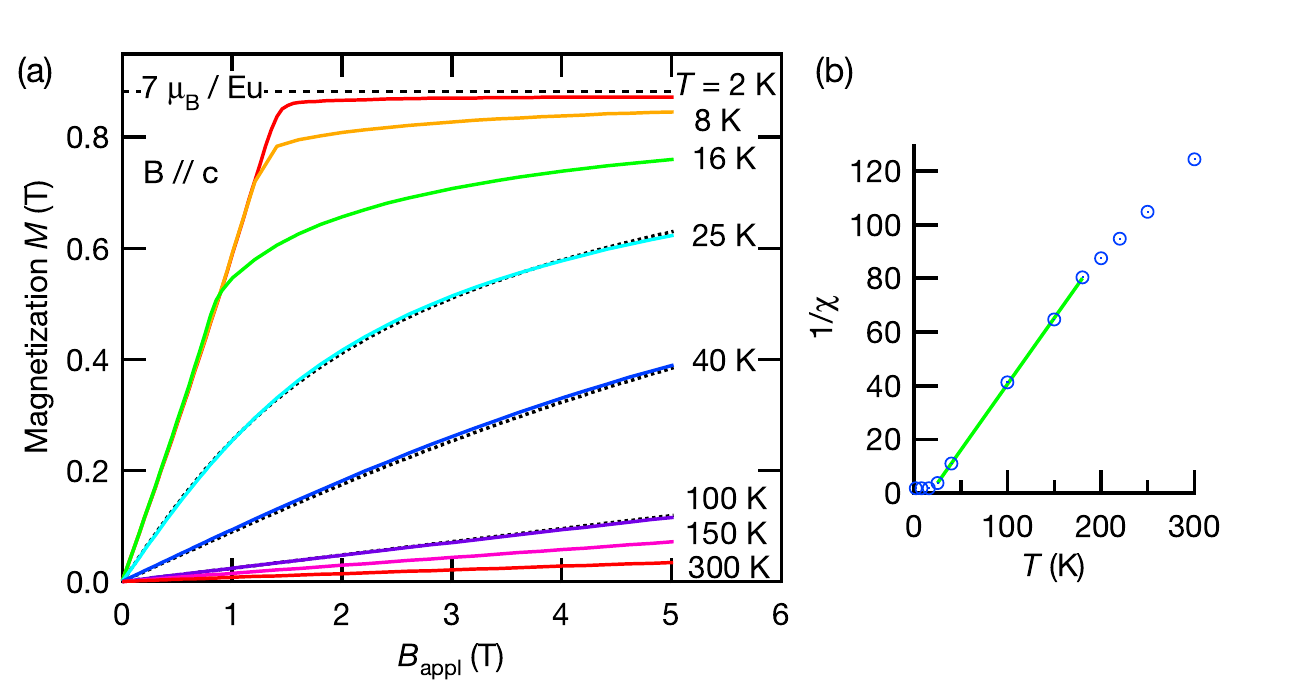}
\end{center}
\caption{(Color online) (a) Magnetization curves of EuFe$_2$As$_2$.  The dotted curves drawn for $T$ = 25, 40 and 100 K are calculated from a molecular-field model (see text). (b) Reciprocal susceptibility 1/$\chi$ as a function of $T$, where $\chi$ is estimated as $\chi$ = d$M$/d$B_{appl}$ at $B_{appl}$ = 0.  The solid line indicates Curie-Weiss behavior below $T_o$.}
\label{MH}
\end{figure}

Figure~\ref{MH}(a) shows magnetization curves for $B_{appl} \parallel c$.  The saturation moment determined at $T$ = 2 K and $B_{appl}$ = 5 T is 6.92 $\mu_B$, close to the theoretical value of 7 $\mu_B$ for the Eu$^{2+}$ ion and consistent with a previous report.\cite{Jiang09NJP}   The magnetization curves in the paramagnetic state of the Eu$^{2+}$ moments can be described very well by using a simple molecular-field model including only the major ferromagnetic interaction.  Namely, the magnetization is given by the equation $M = Ng\mu_BJ$B$_J$($x$) with $x = g\mu_BJH_{eff}/k_BT$, where the effective field is the sum of the applied field and the molecular field: $\mu_oH_{eff} = B_{appl} + AM$.  $B_J$ is the Brillouin function, $J$ = 7/2 for Eu$^{2+}$, and other symbols are as usual.  The dotted curves drawn for $T$ = 25, 40 and 100 K are calculated with $g$ = 2.1 and $A$ = 9.3, which were determined from a fit to the $T$ = 25 K data.  The experimental curves are reproduced satisfactorily.  Figure~\ref{MH}(b) shows the temperature dependence of the reciprocal susceptibility 1/$\chi$, where $\chi$ has been estimated from the magnetization curves as $\chi$ = d$M$/d$B_{appl}$ at $B_{appl}$ = 0.  The susceptibility obeys the Curie-Weiss law $\chi$ = $C$/($T$-$\theta$) below $T_o$ as indicated by a straight line, which gives the effective moment $\mu_{eff}$ = 8.48 $\mu_B$ and the paramagnetic Curie temperature $\theta$ = 17 K.  These values are similar to those previously reported.\cite{Jiang09NJP}  The positive $\theta$ confirms the dominance of the ferromagnetic interaction between the Eu$^{2+}$ moments.

\begin{figure*}[tb]
\begin{center}
\includegraphics[width=17cm]{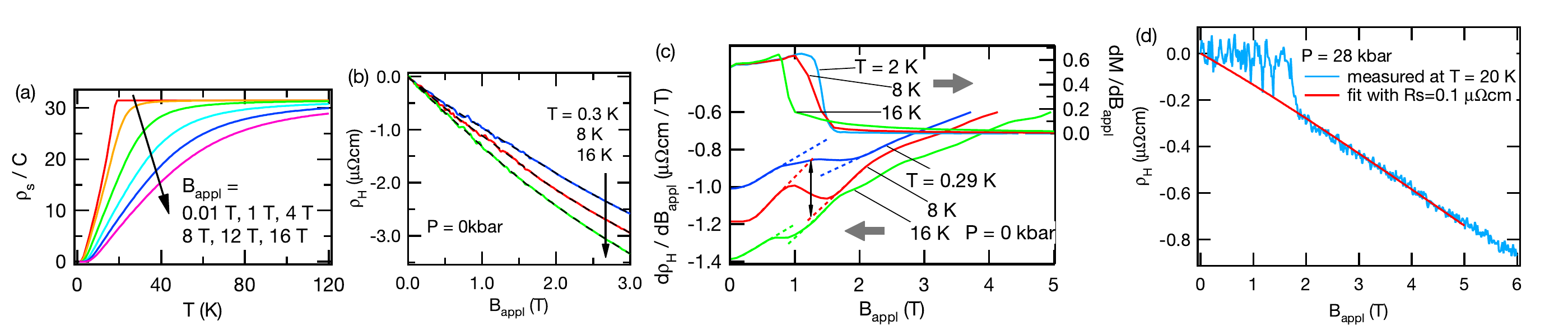}
\end{center}
\caption{(Color online) (a) Simulation of the Eu$^{2+}$ spin-disorder resistivity $\rho_s$ given in the units of $C$.  At $P$ = 0 kbar, $C$ is roughly estimated to be 0.4 $\mu\Omega$cm (see text).  (b) Low-$B_{appl}$ part of $\rho_H$ below $T_N$ at $P$ = 0 kbar.  The dashed curves are spline-function smoothed.  (c) Comparison of $\mathrm{d}\rho_H/\mathrm{d}B$, calculated from the spline-function smoothed curves in (b), and $\mathrm{d}M/\mathrm{d}B$ below $T_N$ at $P$ = 0 kbar.  (d) $\rho_H$ at $T$ = 20 K and $P$ = 28 kbar and a fit with $R_s$.}
\label{sim}
\end{figure*}

We now consider the influence of the Eu$^{2+}$ moments on the transport properties of EuFe$_2$As$_2$.  We start with the resistivity.  The contribution $\rho_s$ of the spin disorder scattering to the resistivity in local-moments systems was previously calculated within the molecular field approximation.\cite{Yamada73JPSJ}  According to the model, $\rho_s = \rho_l + \rho_t$, where $\rho_l = 2C\mathrm{b}^{\prime}(z)$ and $\rho_t = 2C \mathrm{b}(z) (z /2) / (\sinh(z /2))^2$.  The function $\mathrm{b}(z)$ is given by $\mathrm{b}(z) = J \mathrm{B}_J(zJ)$, $\mathrm{b}^{\prime}(z) = (\mathrm{d/d}z)\mathrm{b}(z)$, and $z = x/J$, where $x$ is defined above.  The proportionality factor $C$ depends on the carrier number, effective mass, exchange coupling between the local moments and conduction electrons, but here it is treated as an adjustable parameter.  Figure~\ref{sim}(a) shows simulated $\rho_s$($T$) at various applied fields.  The proportionality factor $C$ at $P$ = 0 kbar may be evaluated as follows.  In the model calculation, $\rho_s(T>T_N)-\rho_s(0)=31.5C$ at zero applied field.  This can be equated with $\rho_o^{\prime} - \rho_o$ = 12.8 $\mu\Omega$cm, where $\rho_o$ is the zero-temperature resistivity extrapolated from the resistivity measured below $T_N$ and  $\rho_o^{\prime}$ is that extrapolated from the resistivity \textit{above} $T_N$.  Thus $C$ is roughly estimated to be 0.4 $\mu\Omega$cm.  To see the appropriateness of this estimation, we have calculated $\rho_s(B_{appl})$ at $T$ = 25 K, and have compared it with the experimental resistivity (see the dotted curve in Fig.~\ref{transport}(b)).  The calculated curve reproduces the observed negative MR at low fields very well, though the positive MR due to the cyclotron motion prevails at high fields.  Using $C$ = 0.4 $\mu\Omega$cm, we can now estimate the ratio of the temperature variation of the Eu$^{2+}$ spin-disorder resistivity to the total measured temperature variation of the resistivity.  In a temperature range between 25 and 100 K at $P$ = 0 kbar, the ratio is about 1\% at $B_{appl}$ = 1 T, increases with $B_{appl}$, and is about 14\% at $B_{appl}$ = 16 T.  Namely, the Eu$^{2+}$ spin-disorder resistivity is not a dominant factor to determine the temperature dependence of the resistivity in this temperature range at $P$ = 0 kbar.  Unfortunately, it is difficult to estimate $C$ at $P$ = 28 kbar, since the resistivity at low temperatures and low fields is inaccessible because of the SC.  However, it is safely assumed that the relative contribution of the Eu$^{2+}$ spin-disorder resistivity at $P$ = 28 kbar is smaller than that at $P$ = 0 kbar.  The reason is that the Fe moments do not order down to zero temperature and hence that the contribution of the Fe spin fluctuations to the resistivity is expected to be larger than that at $P$ = 0 kbar, as is evidenced by the negative MR observed in a wide temperature range.

We now turn to the Hall effect.  The measured Hall effect may contain a contribution from an anomalous Hall effect due to the Eu$^{2+}$ moments.  In order to estimate its size, we follow analyses previously applied to the anomalous Hall effect in rare-earth metals and assume that $\rho_H = R_oB + R_s M$, where $R_o$ and $R_s$ are the ordinary and anomalous Hall coefficients, respectively.\cite{Hurd72}  Note that, since thin samples are usually used for Hall effect measurements, $B$ is practically equal to $B_{appl}$.  It is known that $R_s$ shows a temperature dependence in a magnetically ordered state and that the maximum of $|R_s|$ occurs near the transition temperature.\cite{Hurd72}  On the other hand, $R_s$ in the paramagnetic state is known to be independent of temperature.\cite{Hurd72}  Differentiating the formula, we have $\mathrm{d}\rho_H/\mathrm{d}B = R_o + R_s \mathrm{d}M/\mathrm{d}B$, neglecting a possible slow variation of $R_o$ with $B$.  A large quick change in $\mathrm{d}M/\mathrm{d}B$ is therefore expected to be reflected in $\mathrm{d}\rho_H/\mathrm{d}B$.  We compare $\mathrm{d}M/\mathrm{d}B$ and $\mathrm{d}\rho_H/\mathrm{d}B$ measured below $T_N$ at $P$ = 0 kbar in Fig.~\ref{sim}(c).  For the latter, we have spline-function-smoothed the experimental $\rho_H$ curves  before differentiation [Fig.~\ref{sim}(b)].  Because of the saturation of $M$, $\mathrm{d}M/\mathrm{d}B$ exhibits a sharp drop around 1 T, where $\mathrm{d}\rho_H/\mathrm{d}B$ also shows a step-like change.  Using the $T$ = 8 K data, we estimate that $R_s = \Delta(\mathrm{d}\rho_H/\mathrm{d}B)/\Delta(\mathrm{d}M/\mathrm{d}B) \sim 0.5 \mu\Omega$cm/T.  Since $\mathrm{d}M/\mathrm{d}B \sim 0.6$ at $B_{appl}$ = 0, $R_H = R_o + 0.3 \mu\Omega$cm/T.  Namely, $|R_H| = 1.2 \mu\Omega$cm/T at $T$ = 8 K differs from $|R_o| = 1.5 \mu\Omega$cm/T by about 20\%.  More important is an examination of the Hall effect above $T_N$ in the paramagnetic state of the Fe moments when $T_o$ is suppressed.  Analysis of the Hall effect in the paramagnetic state is usually done by plotting $\rho_H/B$ at various temperatures against $M/B$.  According to the above formula, the plot gives a straight line, from which $R_o$ and $R_s$ can be determined, provided that $R_o$ and $R_s$ are $T$-independent.  However, this method can not be applied to the present case, since $R_o$ is $T$-dependent.  As an alternative, we make use of the fact that, while $M(B_{appl})$ curves at temperatures above but close to $T_N$ are nonlinear, $\rho_H(B_{appl})$ curves at those temperatures at $P$ = 28 kbar are almost linear.  Figure~\ref{sim}(d) shows $\rho_H$ at $T$ = 20 K and $P$ = 28 kbar and a fit to $\rho_H = R_oB + R_s M$, where $R_s$ = 0.1 $\mu\Omega$cm/T has been fixed and the $B$-dependence of $M$ has been assumed to be the same as that observed at $P$ = 0 kbar and $T$ = 25 K.  The fitted curve is close to the lower bound of the scattering of the experimental data points at 5 T, and it is clear that the curve will deviate from the experimental one above 5 T.  We can therefore conclude that $|R_s|$ is significantly smaller than 0.1 $\mu\Omega$cm/T.  If we assume $|R_s|=0.05 \mu\Omega$cm/T, its contribution to $|R_H|$, i.e., $|R_s|M/B$, is $\sim$0.01 $\mu\Omega$cm/T at $T$ = 25 K, which is less than 10\% of $|R_H|$ at $P$ = 28 kbar and $T$ = 20 K.  These analyses indicate that, although the influence of the anomalous Hall effect due to Eu$^{2+}$ may not completely be neglected, the main part of the Hall effect in EuFe$_2$As$_2$ is still due to the ordinary effect.  

In conclusion, we have considered the influence of the Eu$^{2+}$ moments on the electronic transport properties of EuFe$_2$As$_2$ and have found that electron scattering due to the Eu$^{2+}$ moments plays only a minor role in both the resistivity and the Hall effect.


\end{document}